\documentclass{kluwer}    % Specifies the document style.
\newdisplay{guess}{Conjecture}
\def\s|{\left|\!\left|}
\def\d|{\right|\!\right|}
\def\Hcal{{\cal H}}
\def\Ncal{{\cal N}}
\def\bfHcal{{\mbox{\boldmath$\Hcal$}}}
\def\bfnu{{\mbox{\boldmath$\nu$}}}
\def\be{\begin{equation}}
\def\ee{\end{equation}}

\begin{document}                                                                                   
\begin{article}
\begin{opening}         
\title{{\sl Strong Chaos} in N-body problem and {\sl Microcanonical
Thermodynamics} of collisionless self gravitating systems.} 
\author{Piero \surname{Cipriani}\thanks{Permanent address: {\bf C.S.S.} - via E.Nardi, 14/16 -
02047 - Poggio Mirteto (RI) Italia - Electronic address: piero.cipriani@roma1.infn.it 
or cipriani@icra.it}}  
\institute{Dipartimento di Fisica - Universit\`a di Roma {\it "La Sapienza"} - 
P.le A.Moro, 2 - 00185 ROMA (Italia)}
\author{Marco \surname{Pettini}\thanks{Also I.N.F.N., Sezione di Firenze - 
Electronic address: pettini@arcetri.astro.it}}  
\institute{Osservatorio Astrofisico di Arcetri
L.go E.Fermi, 5 - 50125 FIRENZE (Italia)}
\runningauthor{Piero Cipriani \&\ Marco Pettini}
\runningtitle{From Chaotic Dynamics to Thermodynamics of self gravitating systems.}
\date{\today}

\begin{abstract}
The dynamical justifications which lie at the basis of an {\it effective} Statistical
Mechanics for self gravitating systems are formulated, analyzing some among 
the well known obstacles thought to prevent a {\sl rigorous} Statistical treatment. 
It is shown that N-body gravitational systems satisfy a {\sl strong chaos} criterion,
so supporting the assumption of an increasingly uniform spreading of orbits over the constant
energy surface, {\it i.e.}, the asymptotic evolution towards a {\sl microcanonical}
distribution.
We then focus on the necessary conditions for the {\sl equivalence of 
statistical ensembles} and remark that this equivalence is broken for any N-body 
system whose interaction has a range comparable with its spatial extension.
Once realized that the obstacles originate from the long range nature of 
the Newtonian interaction, leading to the non-extensivity of canonical
and grand-canonical thermodynamic potentials, we show that instead 
a suitably generalized {\sl microcanonical ensemble} constitutes an {\it orthode},
{\it i.e.}, a reliable framework for a {\sl correct} Thermodynamics.
Within this setting we use the thermodynamic relations and find consistent
definitions for Entropy (which turns out to be {\it extensive}, despite
the non stable nature of the interaction), Temperature and Heat Capacity. 
Then, a {\sl Second Law-like} criterion is used to select the hierarchy of {\sl secular 
equilibria} describing, for any finite time,
the macroscopic behaviour of self gravitating systems.
\end{abstract}

\end{opening}           

\section{Introduction}
Although is the most {\sl ancient} among the known {\sl fundamental} interactions, 
Gravity seems nevertheless to be the most reluctant to be included 
coherently within the theoretic frameworks describing the behaviour of 
{\sl microscopic} and {\sl macroscopic} many body systems, respectively.
 
Indeed, in addition to the well known difficulties encountered by the attempts
for a coherent formulation of a Quantum Theory of Gravity, there are the less {\sl popular}, 
though equally relevant from a conceptual viewpoint, obstacles preventing a 
{\sl rigorous} Statistical Mechanics of collisionless self gravitating 
N-body systems \cite{Ruelle}.

This is due to the fact that the Hamiltonian function describing the gravitational 
interaction between $N$ point particles of mass $m_i$
\be
H(q,p) = \sum_{i=1}^{N} \frac{{\bf p}_i^2}{2 m_i} \ -\ 
\frac{1}{2} \sum_{i=1}^{N}\sum_{j\neq i} \frac{Gm_im_j}{r_{ij}}, \ \ {\rm where}\  
r_{ij}\equiv \|{\bf q}_j - {\bf q}_i\|
\label{eq:NBHam}
\ee
 does not satisfy the {\it stability} property, which is a prerequisite for a
self consistent Statistical Mechanical description and for the proofs of most 
of the theorems assuring the equivalence of the Statistical ensembles in the so-called
{\sl Thermodynamic limit}.
The lack of ensemble equivalence for N-body self gravitating systems (SGS),
leads naturally to the issue of determining
which, if any, among the many ensembles complies with the {\it orthodic} 
condition \cite{GallaMS}.

Notwithstanding the hindrances for a Statistical Mechanics of gravitational
systems are known since many decades, many efforts have been devoted to overcome
them, in the attempt to formulate, at least, an {\it effective} Statistical
description of collisionless N-body SGS \cite{Chandrasekhar,Padmanabhan,Saslaw,BT}.

The astrophysical motivations for doing so are well evident looking at the
otherwise inexplicable morphological, structural and dynamical similarities 
among {\sl hot} stellar systems observed in the sky. 
The {\sl qualitative} homogeneities of the luminosity curves and of velocity 
dispersion profiles, though with obvious {\sl quantitative} 
differences, suggest unavoidably the existence of one or more relaxation processes
leading globular clusters and elliptical galaxies towards an equilibrium state
which is qualitatively independent on the global parameters of the systems and
on the class of initial conditions of these astronomical objects at the formation era.

Due to the peculiar nature of Gravity, the relaxation mechanisms ({\tt e.g.}, binary
{\it collisions}) at work in usual
{\sl laboratory} many body systems turn out to be unable to justify such a state
of ({\it quasi}) equilibrium, because the corresponding {\sl relaxation time} predicted is
comparable to he Hubble time for globular clusters, and several orders of magnitude larger
for elliptical galaxies.

More than thirty years ago, Lynden-Bell, Saslaw and others \cite{LB_VR,SaslawVR}
first realized that the infinite range of Newtonian interaction could support
relaxation mechanisms related to {\sl collective effects}, in turn due to 
global, almost coherent, oscillations of the gravitational field, able to
change appreciably the energies of the particles ({\it i.e.}, the stars).
However, the general relevance of such a mechanism has been revised and even
criticized to some extent, \cite{Shu-Madsen,Gerhard-Feix} and it is 
by now agreed that, though useful to understand some aspects of the initial
global evolution of gravitational systems, it alone cannot by sufficient to 
explain the {\sl universal} light and velocity profiles observed.\\

The lack of a satisfactory Statistical Mechanics for SGS made problematic obviously
any attempt to formulate a Thermodynamics for them. Among the various {\sl peculiarities}
of gravitational interaction (discussed in the sections below), those implying
the {\it non extensivity} of (dynamic and) thermodynamic potentials constitute
the main obstacles to the definition of thermodynamic quantities and, even more,
to derive {\sl standard} relationships among them.

A major piece of evidence of the very peculiar Thermodynamics proper to
SGS is known since many decades to astronomers, but it went to know a wide 
{\sl popularity} in the Physics community only when Thirring {\sl rediscovered} this apparent
{\it paradox} \cite{Thirring}, which was already investigated rather
deeply some years before by Antonov \cite{Antonov} and by Lynden-Bell and Wood
\cite{LB&W}. The {\it problematic issue} emerges when we recall from one side the well known 
(to astronomers) property of (bound) SGS of having {\sl negative heat capacities}, and from the
other the expression of the heat capacity as it can be easily derived
within the Canonical ensemble (see, {\tt e.g.}, \cite{Huang}, \S 8.2):
\be
\frac{\partial U}{\partial\beta} + \langle(H-U)^2\rangle = 0\ ,
\label{eq:dudbeta}
\ee
where $U\equiv\langle H\rangle$ is the average value of the energy of the
system in the canonical ensemble, $\beta=1/k_B T$ is the (inverse) temperature 
parameter of the canonical distribution and $k_B$ is the Boltzmann constant. 
Written in terms of the heat capacity $C_V$, equation (\ref{eq:dudbeta}) reads
\be
C_V = \frac{1}{k_B T^2} \left( \langle H^2\rangle - \langle H\rangle^2 \right)\ \geq\ 0\ .
\label{eq:CV}
\ee

In the recent years, however, it has been realized that such a paradox is not
a peculiarity exclusive to SGS, but it is instead a common features of {\it small systems},
{\it i.e.}, systems whose extension is smaller than (or comparable to) the range
of the governing interaction \cite{smallsystems,LB&LB,Chomaz00}. For these
systems it is immediately realized that {\it extensivity} property breaks down.
However these systems possess a well defined, though in some respects peculiar,
thermodynamical behaviour, and so it turns out as well that such a property 
is not necessarily relevant to allow for a description of the macroscopic, 
{\sl quasi-thermodynamic} behaviour of these systems, as discussed
in more detail below, where we will analyze also why this feature obstructed
to some extent the formulation of a Thermodynamic framework for SGS.

An immediate consequence of the lack of extensivity is the breakdown of the 
{\sl ensemble equivalence}, which is another milestone of the {\sl rigorous}
statistical mechanical foundation of the thermodynamic formalism. However, the 
phenomenology of non extensive systems is currently investigated from theoretical,
numerical and also experimental viewpoints, and many studies nowadays focuse
on the dynamical, statistical and thermodynamic properties of Hamiltonian systems 
characterized by ensemble {\sf in}equivalence \cite{smallsystems,Chomaz00,Dauxois00}.\\

According to the considerations above, the attempt to formulate a {\sl coherent} 
thermodynamic description of SGS appear to be feasible, starting from the basic 
assumptions which lie at the grounds of the Statistical description of many degrees 
of freedom dynamical systems, though paying due attention to every possible source of
internal inconsistency.

Then, going back to Dynamics and Statistical Mechanics, we first investigate the requisites
to be fulfilled by the Dynamics, in order to justify a statistical description of 
macroscopic properties of N-body systems. This goal is accomplished using some
concepts of Ergodic theory of (many degrees of freedom) dynamical systems,
and in particular we adopt the {\sl Geometric Description} of Dynamics,
which allows to relate the chaotic properties of motions to the geometrical
and topological properties of the dynamical manifolds associated to the 
system, see, {\tt e.g.}, \cite{Marco93,CPTD,MarcoLapoLivi,PSS98,PRL98,MarcoLapoCohen2000}.

Once found convincing indications that a
self consistent {\it effective} Statistical Mechanics of SGS can be formulated\footnote{Though
with some peculiarities which are the statistical counterparts of the dynamical
unique features of gravitational interaction.}, we then face the issue of the
ensemble inequivalence and argue that the Classical {\sl Microcanonical ensemble} alone
is suitable for a reliable and coherent description of the Thermodynamics of
non extensive systems like gravitational N-body ones. The criteria according to which
the Microcanonical distribution over the hypersurface
$\Sigma_{\cal E}$ of the full $6N$-dimensional phase space $\Gamma$,
 acquires such a privileged role over other
classical ensembles are analyzed and discussed, and it is shown how such a choice
allows to derive a number of {\sl physically reasonable} results on the macroscopic
properties of N-body gravitating systems as well as unambiguous (re-)definitions of 
thermodynamic quantities and relations.

As a by-product of the above analysis, we also derive and discuss some among the many non trivial 
relationships existing between the statistical properties\footnote{As, for example,
ergodicity of dynamics in $\Gamma$ space or even more strong features, like
{\it mixing}.} of dynamics over the constant energy
hypersurface $\Sigma_{\cal E}$, the consequent statistical mechanical description,
and the customary {\it reduced} description in terms of a {\sl distribution
function} ({\sc df}) over the single particle six dimensional $\mu$-phase space, depending only, 
according to Jeans theorem \cite{BT}, on the isolating integrals of motion admitted 
by the {\sl smoothed} potential in the three dimensional configuration space. 
This theorem indeed is at the
grounds of most of the works devoted to the construction of self-consistent models of
stellar systems over the last four decades, and only recently it has been realized that
its relevance for stellar dynamical modelling, at least in the customary {\sl strong formulation} 
sketched above, has been somewhat overemphasized \cite{JeansThm}.
We outline here the conditions to be satisfied by the full dynamics in
$\Gamma_{6N}$-space in order to allow for the {\sl reduction process} leading to a 
hydrodynamic (or kinetic) description in terms of a distribution function over the $\mu$-space.
This represents a first step along the path of a critical reconsideration of the use
of the single particle {\sc df} as a tool to derive the equilibrium properties of
stellar systems through {\sl maximum entropy} like arguments based on it 
\cite{MaxEntDF}.
A deeper discussion of this topic is however beyond the scope of this paper
and will be the subject of forthcoming works.\\

The paper is organised as follows: in section \ref{sec:Dynamics} we study the chaotic
properties of many degrees of freedom systems, with particular emphasis
on the features of the dynamics of SGS and recall the usuals indicators adopted to
quantify the degree of stochasticity of orbits in phase space.  
In order to relate Dynamics to the structure of the ambient space, we adopt the
terminology and the tools of the Geometric Description of Dynamics, which provides
an elegant and effective framework to characterize the intrinsic properties
of dynamical systems. 

To support our argumentations we recall some of the results obtained within
this framework and reinterpret them in the perspective of the present
work, devoted to propose thermodynamical description of SGS, justified on the grounds
of the statistical features of dynamics and of the geometric properties of the dynamical
manifolds.
In section \ref{sec:FromDynTOSM} we briefly introduce some concepts from the Ergodic
theory of Dynamical Systems and use them to classify the statistical properties of the
flow associated to the N-body dynamics, arguing for the possibility of a 
statistical description of their macroscopic properties on any finite scale of times. 
In section \ref{sec:TD} we review the issues related to the non equivalence of ensembles, then
we discuss why the classical Microcanonical ensemble is the only legitimate to describe
coherently the {\it "Thermodynamics"} of SGS; we then present some consequences of
the thermodynamic relations derived, analyzing also the various inconsistencies encountered
attempting to describe the system within the {\sl canonical} ensemble. 
Section \ref{sec:Concl} presents a brief discussion
of the results, the {\sl peculiarities} of the Thermodynamics of systems without
{\sl Thermodynamic limit}, a heuristic discussion on the concept of {\sl finite time
equilibrium} and concludes outlining the work in progress along the line here reported, 
possible extensions of the formalism and the improvements to the results presented
which can be obtained with further studies and more extensive numerical work.

\section{Chaotic behaviour and Ergodic properties of Dynamics of gravitational 
N-body systems. \label{sec:Dynamics}}  
Since the first {\sl rediscoveries}\footnote{Most of the conceptual
implications of dynamical instability as well as the realization that such a phenomenology
should be the rule rather than the exception in the behavior of nonlinear
({\it i.e.}, generic) dynamical systems were indeed already present in the
Poincar\'e work \cite{Poinc} more than a century ago!} of the ubiquity of Chaos in nonlinear
dynamical systems in the 60's, \cite{HH}, it was realized that unpredictability implied
by the exponential sensitivity to small changes in initial conditions could be the
ingredient needed to justify a more or less {\sl physical} reformulation of the
{\it ergodic hypotesis}, able to explain the approach to a stationary equilibrium state
of most non integrable many degrees of freedom (MDoF) Dynamical Systems. To make a long story
short, it is by now generally agreed that the presence of Chaos entails a system to forget
the detailed memory of initial conditions, though some {\it imprinting} related to
few global constraints or {\sl quasi conserved} quantities could survive during the
dynamical evolution, leading to a slow decay in time of specific correlation functions.
The sufficient conditions implying irreversible behaviour and then able to justify
a statistical mechanical description of macroscopic properties 
have been mathematically demonstrated only for very special
(and rather abstract) classes of models, however there is a widespread agreement that,
from a physical viewpoint, most MDoF nonlinear dynamical systems 
are indeed {\sl chaotic enough} to relax at least to a stationary (meta-) equilibrium state.

\subsection{Chaos, Ergodicity and Mixing.}
For few dimensional systems Chaos is not sufficient to guarantee the efficiency of the
approach to equilibrium, because diffusion in phase space can be inhibited by the fractal
structure of deformed (or partially destroyed) invariant surfaces describing the
{\sl integrable limit}, {\it i.e.}, by the {\it Cantori} originated by the disappearance
of KAM-tori \cite{KAM}. Even above the KAM threshold, the exponentially long times associated to
the Nekhoroshev regime\footnote{During which the actions are almost exactly
conserved and dynamical trajectories are trapped between confining surfaces.}\cite{Nekh}, the 
relaxation processes are practically {\sl frozen} and from a physical viewpoint the
approach to equilibrium will not take place at any finite time \cite{Galgani}.

However, both KAM and Nekhoroshev thresholds depends strongly on the dimensionality of
phase space\footnote{Throughout the paper we indicate with ${\cal N}=N\cdot d$, the number
of degrees of freedom of a dynamical system, $N$ being the number of particles and $d$ the 
dimensionality of the single particle configuration space. For Hamiltonian systems then the
phase space $\Gamma$ has dimension $2{\cal N}$ and the constant energy hypersurface $\Sigma_{\cal E}$
has clearly dimension $2{\cal N}-1$.}, decreasing roughly exponentially with the number 
of degrees of freedom. Together with the Poincar\'e-Fermi \cite{Fermi} theorem on the non existence of
analytic integrals for nonlinear systems with three or more degrees of freedom, the vanishing
of the integrability or quasi integrability thresholds, suggests that MDoF chaotic systems
should attain the {\it asymptotic equilibrium states}.

However, from a physical viewpoint, it is also important that the approach to such states
be fast enough, that is, it is required that the time averages of relevant dynamical
macroscopic quantities converge to corresponding phase averages in finite times and not only
asymptotically. Rephrased in terms of Statistical theory of Dynamical Systems, 
this amounts to say that dynamics leads any finite volume in phase space to spread out 
uniformly over the entire region accessible to motion. Formally, in addition to the
{\it ergodic} property:
\be
\lim_{T\rightarrow\infty} \frac{1}{T} \int_0^T F[{\bf q}(t),{\bf p}(t)] dt\ =\ 
\frac{1}{{\cal M}(\Gamma)} \int_\Gamma F({\bf q},{\bf p}) \mu(d{\bf x})\ ,
\label{eq:ergodicity}
\ee
the dynamical evolution must also satisfy the stronger condition known as {\it mixing}:
\be
\lim_{T\rightarrow\infty} \int_\Gamma F[{\bf q}(t),{\bf p}(t)] G[{\bf q},{\bf p}] \mu(d{\bf x})\ 
=\ \left(\int_\Gamma F({\bf q},{\bf p}) \mu(d{\bf x})\right)\cdot
\left(\int_\Gamma G({\bf q},{\bf p}) \mu(d{\bf x})\right)\ ,
\label{eq:mixing}
\ee
where $F$ and $G$ are arbitrary functions of dynamical variables
and $\mu(d{\bf x})$ is the invariant measure over $\Gamma$\footnote{What is relevant
from the point of view of Physics is that equation \ref{eq:mixing} be satisfied at least by
macroscopic {\sl relevant} functions $F$ and not necessarily for {\it any} pair $F$ and $G$, as instead
is required in the mathematical definition of {\sl mixing}.
In the case of Hamiltonian systems the invariant measure is the Liouville measure.}.

As above, the {\sl mixing} property has been rigorously proved only for restricted classes
of abstract dynamical systems, {\tt e.g.},\cite{KAM}, though many theoretical
and numerical works \cite{Trelax} now support strongly the belief that
even more realistic models of many body physical systems possess {\sl good} statistical
properties and approach the equilibrium state on physically reasonable time scales;
moreover, these statistical relaxation times bear close correlations with the
orbit diffusion rates in phase space, and also with Lyapunov exponents describing the degree
of chaoticity.

The above evidences have been given a formal expression recently through the formulation
of the so-called {\sl Chaotic Hypotesis} \cite{GC95}, which puts on the same level, for what
concerns the statistical description of macroscopic properties, chaotic many degrees of freedom
dynamical systems and Anosov systems, which are, among the {\sl abstract} models, those characterized
by the strongest statistical properties (together with {\sl Axiom A} and {\sl K-systems}, 
see, {\tt e.g.}, \cite{LebPen})\footnote{These models, though maximally chaotic, are,
for what concerns the evolution of their statistical properties, almost as simple as
 integrable systems  (\cite{CPTD},chapters 4$\div$6).}.

Before to proceed further, it is necessary to remark that all the above discussion about
the KAM and Nekhoroshev thresholds is of fundamental relevance to single out the sources of Chaos
in MDoF dynamical systems and for the understanding of the relationships between the
instability of the trajectories and the modifications occurring in the structure
of phase space. Nevertheless, most of the tools and concepts developed within that framework
are of little use for the characterization of the dynamical properties of
SGS in an arbitrary state, because there is nothing indicating the existence 
of any integrable limit for gravitational N-body systems; {\it i.e.}, for them,
there is no hope to undertake a sensible perturbative approach to study the onset of Chaos,
at least as long as $\ln{N}\geq 1$. This amounts to say that the investigation of dynamical
instability in SGS must resort to methods and tools pertaining more to the theory of strongly
chaotic dynamical systems rather than to Perturbation Theory, whose approaches are suited
for nearly integrable systems.  

\subsection{Geometry of Dynamics and Chaos.}
Within the class of systems for which the {\sl mixing} property has been rigorously proved there
are the {\sl geodesic flows on manifolds of constant negative curvature} \cite{Anosov}, which were
studied in the 30's by Hedlund and Hopf. The first to realize the relevance of such a
line of research to Statistical Physics was Krylov \cite{Krylov}, which studied the consequences
of the instability of the geodesic flow associated to the dynamics of an hard sphere gas, predicting
that the instability leads the system to macroscopically approach the equilibrium state, with
a relaxation time which agrees rather well with more recent estimates.

The formal apparatus for the geometric transcription of the dynamics of natural systems was developed
independently by many authors and found a beatiful presentation in the Synge's article
{\it "On the Geometry of Dynamics"} \cite{Synge}. Although completely unaware of the relevance
of exponential instability, he there discussed extensively the stability of geodesic flows
of dynamical interest.\\

The fundamental argument able to relate geodesic instability and the entailed chaotic dynamics
from one side, to the fulfillment of the {\sl mixing} property on the other, goes through the proof
that the dynamical manifold is characterised by almost everywhere negative sectional curvatures.
If this is the case, then the {\sl Instability time}, as measured by a suitable generalization of the
(inverse) Lyapunov exponent, is closely related to the {Relaxation time}, giving the scale over which
correlations decay and the memory of the initial state is lost. Then the point has been moved from
a direct proof of mixing to the study of the sign of the sectional curvatures of the manifold.

We refer for the details of the geometrical transcription of dynamics of Hamiltonian and Lagrangian
systems to \cite{MarcoLapoCohen2000,PSS98} and references therein, and simply recall that the tool for
investigating the stability of the geodesic flow are the {\sl Jacobi--Levi-Civita} equations
for geodesic deviation, which, in local coordinates over a suitable manifold read:
\be
{{\nabla} \over {ds}}\left({{\nabla \delta q^a} \over {ds}}\right)
 + {\cal H}^a{}_c \delta q^c = 0 \ , \qquad
(a=1,\ldots,{\cal N}) 
\label{eq:EDG}
\ee
where ${\delta{\bf q}}$ represents the perturbation to the flow, whose (possibly) exponential 
growth is the main signature of instability and depends on the {\sl stability  tensor} 
${\cal H}^a{}_c$, defined 
{\it along} a given geodesic (with unitary tangent vector ${\bf u}$),
 through the Riemann curvature tensor, $R^a{}_{bcd}$ of the manifold (with metric $ g_{ab}), $ by
\be
{\cal H}^a{}_c \doteq  R^a{}_{bcd} u^b u^d\ ;
\ee
and $\nabla/{ds}$ is the total (covariant) derivative along the geodesic.
From equation (\ref{eq:EDG}), an effective ordinary differential equation for the norm of the 
perturbation can be derived. Defining an unit vector ${\bfnu}$ on the unitary
tangent bundle of the manifold {\sf M}, ${\sf T}_{\bf q}{\sf M}$, such that $\delta{\bf q}=z{\bfnu}$, it
is straightforward, \cite{PSS98}, to show that the magnitude of the 
perturbation, $z$, evolves according to:
\be
{{d^2z}\over{ds^2}} = \left( -{\cal H}_{ac} \nu^a \nu^c + \s| 
{{\nabla{\bfnu}}\over{ds}}\d| ^2 \right) z 
\label{eq:norma1}
\ee
which is still an {\sl exact} equation containing all the required informations about
the overall instability of the flow. The simplification obtained going from equation(\ref{eq:EDG}) 
to equation(\ref{eq:norma1}) is rather illusory. Indeed, even if we reduced a system of ${\cal N}$ 
differential equations containing covariant derivatives apparently to just one ordinary differential
equation, the computational task has been made only a little easier.
The problem is that equation (\ref{eq:norma1}) still contains the full stability tensor, 
whose explicit computation requires the knowledge of the ${\cal O}({\cal N}^4)$ components of the 
Riemann tensor. In some particular cases it happens that the manifold is such that the simplification
obtained is actually decisive \cite{SymSpace}, but as expected, this is not true for most
geometric transcriptions of realistic models of physical systems.

For MDoF systems, however, in the hypotesis, to be verified {\it a posteriori}, that the dynamics
is {\sl far enough from complete integrability}\footnote{What is required actually is much less: 
it is sufficient that the evolution of the many degrees of freedom is not coherent, that is, 
that there is a spectrum of frequencies wide and far enough from resonances, so that collective 
periodic oscillations are avoided. In other terms, a good {\sl phase mixing} is sufficient to prevent
the persistence of phase correlations among the motions of individual particles; and we recall
that phase mixing occurs in integrable systems with MDoF as well! From the geometrical viewpoint, 
this is enough to imply that the ${\cal N}-1$ non vanishing sectional curvatures are locally almost 
uncorrelated each other.}, it is possible to obtain a {\sl closure} of equation(\ref{eq:norma1}), 
which then reduces to the approximate form:
\be
\frac{d^2z}{ds^2} + {\cal K}_{R_{\bf u}} [{\bf q}(s)]\, z\ \cong\ 0\ ;
\label{eq:norma2} 
\ee
where 
\be
{\cal K}_{R_{\bf u}} [{\bf q}(s)]\doteq \frac{\Hcal^a{}_a}{\Ncal-1}  
\equiv \frac{{\sf Tr}\,\bfHcal}{\Ncal-1} \  ;
\label{eq:curv}
\ee
is nothing else than the {\sl Ricci curvature} in the direction of the geodesic, equal also to
\be
{\cal K}_{R_{\bf u}} [{\bf q}(s)] =  \frac{R_{ab} u^a u^b}{\Ncal-1} \doteq
 \frac{{\sf Ric} (\bf u)}{\Ncal-1}\ ,
\ee
where $R_{ac}\doteq R^b{}_{abc}\equiv g^{bd}R_{abcd}$ is the Ricci tensor.\\

Referring to the above cited bibliography for a detailed discussion, we simply observe here how,
from the effective equation (\ref{eq:norma2}) it is easy to understand the classical results
on the {\it chaotic} properties of geodesic flows on manifolds of negative curvature: if the
Ricci curvature per degree of freedom ${\cal K}_{R_{\bf u}} [{\bf q}(s)]$ is everywhere negative
along a geodesic, then the geodesic flow is exponentially unstable. So, a curvature everywhere
negative (that is, always, along a geodesic) entails an exponential instability of the flow. 

But this is not the only mechanism leading to Chaos. It is indeed possible to have an exponential
growth of the solution of equation (\ref{eq:norma2}) also when the sign of the
Ricci curvature along a geodesic is fluctuating and even for everywhere positive curvature. 

\subsubsection{Sources of instability in positive curvature manifolds.}

Two further sources of instability are indeed possible for equations of the form (\ref{eq:norma2}): 
\begin{itemize}
\item the onset of exponentially unstable solutions can originate (when ${\cal K}_R > 0$) from
the mechanism of {\sl parametric resonance}.
This phenomenon, well known in the linear case (see, {\tt e.g.}, \cite{LL1}), occurs for example in 
the generalization of Mathieu-like equations,
\be
\ddot x + \Omega^2(t) x = 0\ ,\ \ \ \ \Omega^2(t) = \Omega_o^2 [1+f(t)]\ ,\ \ \langle f\rangle_t =0\ ,
\label{eq:ParRes}
\ee
 when the frequency $\Omega^2(t)$ undergoes a periodic or multiperiodic modulation whose frequencies
satisfy suitable resonance conditions with the unperturbed average frequency, $\Omega_o$. 
If the behaviour of parametric resonance in linear systems is rather elementary, when this phenomenon
couples with even small nonlinearities present in the system, the outcome can be very rich and 
complex \cite{CPNcim00}. 
The interest in this mechanism resides in the fact that it appears very often when a phenomenological
modellization of complex systems is attempted through a reduction process leading to the 
formulation of a system of {\sl effective equations}, alike equation (\ref{eq:norma2}) above.
It has been shown ({\tt e.g.}, \cite{PSS98,PRL98,MarcoLapoCohen2000}) that the occurrence
of positive curvatures is the rule, rather than the exception, for most geometrical transcription
of MDoF dynamical systems. As it is known, nearly all nonlinear systems exhibit nevertheless
chaotic behaviours for almost all initial conditions. As every system is characterized by one
or more intrinsic periodicities, often related to some kind of collective behaviour\footnote{
As examples ranging from Solid State Physics to Astrophysics, we mention the resonances occurring
with the lower modes frequencies in a slightly anharmonic crystal, which modulate the dynamics
on the largest possible spatial scales, and the overall quasi periodicity of the gravitational field
inside a stellar systems undergoing the Violent Relaxation phase, which modulate the background
field of existing condensed substructures.
},
it appears reasonable to investigate the corrispondences between the parameter space and the
astonishingly rich behaviour disclosed by evolutionary laws apparently so simple.
In order to give a taste of the various phenomenologies displayed from an equation
with only 1.5 degrees of freedom, we report in figure \ref{fig:NLPR}, the behaviour of different 
solutions of a single equation, obtained adding small nonlinearities to (\ref{eq:ParRes}) and
whose two parameters have been varied slightly to obtain
the different patterns shown in the panels.
\begin{figure} 
%\vspace{6cm}
% Syntax:  \seteps{<position>}{<width>}{<height>}{<path+filename>}
%     Requires "\input seteps" at the beginning of your file.
%     <position> is horizontal offset from the left margin (can be negative).
% Optional:  <path> (use / instead of \), specifies path of TeX file if not supplied.
% Example:  \seteps{5cm}{1in}{2in}{c:/mysubdir/mypic.eps}
%\seteps{0cm}{12cm}{10cm}{1GDL.eps}
\caption[]{Parametric Resonance in a one degree of freedom nonlinear Lagrangian system with time-dependent
potential. Each panel shows the plot of the solution of a single equation with a slightly
different choice for the values of the only two free parameters. 
The behaviour of solutions of the first two panels recall
those of the underlying linear model, with the typical damped and resonant evolution, respectively. 
Other choices of the parameters values originate either quasi periodic solutions, intermittent or
steadily {\it chaotic} patterns.}
\label{fig:NLPR}
\end{figure}
\item
The second mechanism able to lead to exponential instability of the flow is known as
{\sl Stochastic Resonance} and arises instead when the modulation of the frequency is due to 
stochastic fluctuations around its average value \cite{VanKampen}, and is typical of either non 
completely isolated systems, where fluctuations arise from the environmental influence, or, as for
parametric resonance, in the case of {\sl effective equations}, describing the evolution of 
global or collective variables. Now the modulation is not described through deterministic
modellization, but instead originates from the fact that the exact values of the parameters 
(and in particular of the average frequency) are determined by the combined effects of all the 
{\sl microscopic} constituents of the system and are then influenced by the noise due to finite 
size effects.
\end{itemize}
 
As a general rule (though of a rather heuristic nature), parametric resonance is responsible of the
unstable behaviour of few dimensional systems as well as of the dynamical instability occurring
in MDoF not very far from integrability and with an overall intrinsic underlying periodicity, 
whereas, as stated before, stochastic resonance is the typical mechanism promoting the
development of Chaos in strongly non integrable MDoF systems; its description  becomes more and
more accurate as the number of degrees of freedom increases.

It is obviously possible that the two mechanisms operate together, even if in this case it turns out very 
difficult to separate the effects and even to understand the conditions under which they 
enhance or damp each other.
What is more plausible is that there are regimes in a given model in which either of the two mechanisms
dominate over the other. If this is the case, heuristic argumentations suggest that in the intermediate
range the two effects substantially do not interact, otherwise a discontinuous transition should 
be observed\footnote{
As the two mechanisms cause the onset of instability exciting in a very different way the frequency
space (strongly localized the first, with a rather narrow distribution the second), we expect
the effect of the overlap to be in generic cases negligible.}.

From the above discussion we expect that the mechanism leading SGS to Chaos to be more closely related
to Stochastic rather than parametric resonance, and we will see below that this is indeed the case.

\subsubsection{Geometric properties of the manifold and analytic computation of the maximal
Lyapunov exponent.}

In recent years the geometrical description of dynamics has been applied to the investigations
of the dynamical behaviour of a large class of Hamiltonian and Lagrangian systems either with
few or many degrees of freedom, (see, {\tt e.g.}, \cite{Marco93,CPTD,PSS98,PRL98}
and for a comprehensive review and an updated bibliography, \cite{MarcoLapoCohen2000}). 

The {\it GeometroDynamical Approach} (GDA) applied to Hamiltonian MDoF
systems allowed to improve our understanding on many aspects of the characterization, quantification and
sources of chaotic behaviour.
Among the most remarkable, from the point of view of a concise synthetic characterization
of the dynamic instability, it has been possible to use the GDA, in conjunction with
the theory of Stochastic Differential Equations \cite{VanKampen}, 
to derive an effective analytic formula for the maximal Lyapunov exponent 
\cite{MarcoLapoLivi}, which is obtained simply evaluating average values, fluctuations and
autocorrelation time of the {\sl frequency} of the effective equation (\ref{eq:norma2}). 

If we indicate with $k_R\doteq\langle {\cal K}_{R_{\bf u}} [{\bf q}(s)]\rangle$ 
the average (positive) value of the Ricci curvature per degree of freedom, 
with $\tilde{\sigma^2_{k_R}} \doteq
{\cal N} {\sigma^2_{k_R}}$ the rescaled variance of its fluctuations\footnote{
For a discussion of the ${\cal N}$ rescaling of the variance, see \cite{MarcoLapoLivi}.}
 and with
$\tau_{k_R}$ the autocorrelation time of the stochastic process describing the time evolution
of ${\cal K}_{R_{\bf u}} [{\bf q}(s)]$ along any given geodesic, it is possible to
analytically compute the average exponential growth rate of the perturbation according
to the following formula
\be
\lambda_1 = \sqrt{\frac{k_R}{3}} \left( g^{1/3} - g^{-1/3}\right)\ ,
\label{eq:LCN}
\ee
where
\be
g\doteq \frac{1+\sqrt{1+\eta^2}}{\eta}
\ee
and
\be
\eta \doteq \frac{8\sqrt{3}}{9} \frac{k_R^{3/2}}{\tilde{\sigma^2_{k_R}} \tau_{k_R}}\ .
\ee
The importance of equation (\ref{eq:LCN}) relies also on the fact that for most many body systems
the quantities entering it can be computed using either the standard time averages
along numerical trajectories or the {\sl statistical} phase averages over the canonical
or microcanonical distribution. The agreement between the results obtained using either of
the above procedures is rather satisfactory \cite{CPTD,MarcoLapoLivi}, and numerical simulations
along trajectories allow also to check the reliability of the assumptions at the grounds of the
use of the theory of stochastic differential equations, first of all the gaussian distribution
of the values of the curvature.

In addition to this, the GDA helps to improve the conceptual understanding of the onset
of instability in MDoF Hamiltonian systems, showing that, even for the most chaotic N-body
systems, the sources of instability rely mostly on the stochastic fluctuations of the
curvature rather than on the occurrence of negative values. In particular, for what
concerns SGS, it has become clear that dynamical instability of trajectories, originates mainly from
{\sl binary encounters}, and that the latter entail {\it positive fluctuations of the curvature}:
the strong chaoticity of N-body systems is mainly due just to the huge amplitude of 
curvature's fluctuations around the average value, to be compared with what happens in
other MDoF Hamiltonian systems, where fluctuations become of the order of the average values only
in the strongly chaotic limit.

Moreover, the GDA singles out clearly the different
regimes accompanying the approach to equilibrium, as the transition to a {\sl less unstable}
regime when virial equilibrium has been attained and the violent relaxation process accomplished;
and many other phenomena operating on longer timescales and finer spatial extensions.
The details of these issue have been
partially discussed in \cite{PSS98} and will be analysed in a deeper detail elsewhere.

\subsection{Strong Stochasticity threshold and Chaotic Hypotesis.}
To the main issue of this paper what is relevant of the above discussion is the evidence of
a very strong chaoticity in gravitational N-body systems. Without going into formal
details, we present here the results of the application of the above formalism to SGS, 
and compare them with analogous outcomes related to a chain of 
anharmonic oscillators\footnote{The well known Fermi-Pasta-Ulam quartic, or {\sf FPU-}$\beta$, model
\cite{Fermi}.}, 
in order to characterize the degree of chaoticity present in 
N-body dynamics. As anticipated above, in MDoF nonlinear systems there are in general various dynamical
regimes. Starting from quasi integrable behaviour for small nonlinearity (or at enough low
energy), a complete hierarchy of chaotic regimes is encountered as the departure from
integrability increases. The KAM and Nekhoroshev thresholds are very near the {\it zero}
of the perturbation parameter, $\varepsilon$, as soon as the number of degrees of freedom is greater than
a few tens, but  nevertheless, in several different Hamiltonian systems it is possible
to single out a further physically meaningful threshold, separating two different regimes
of stochasticity. This dynamical transition has its statistical counterpart in a rather
abrupt change in the behaviour of the relaxation time of macroscopic quantities.\\ 

Briefly, below such a threshold, let it be $\varepsilon_c$,
the rate of increase of the maximal Lyapunov exponent, $\lambda_1(\varepsilon)$,
 is described by a relatively steep (model dependent) power law and the relaxation times 
are very long, depending sensitively on the perturbation parameter. 
Above this threshold, the growth rate of $\lambda_1(\varepsilon)$ slows down and the 
relaxation times become much shorter and nearly independent of the value of $\varepsilon$
(as long as $\varepsilon > \varepsilon_c$) \cite{Trelax}.

A very interesting and illuminating feature of the scaling law $\lambda_1(\varepsilon)$ above
$\varepsilon_c$\footnote{The so called {\sl strong stochasticity threshold}, {\sf SST}, 
\cite{Marco93,Trelax}.} is such that the dimensionless quantity $\gamma_1\doteq\lambda_1\cdot t_D$,
introduced in \cite{PSS98}, turns out to be virtually independent on the
perturbation parameter. This because, above the {\sf SST}, the maximal Lyapunov exponent scales
exactly as the reciprocal of the dynamical time scale. This evidence implies that the 
diffusion of orbits in phase space proceeds at the fastest rate allowed by the dynamics. 
This fact has been noticed and briefly discussed in \cite{PSS98}, suggests a very elegant scenario
for the mechanisms governing the onset and the development of Chaos, and gives a convincing
picture of the modifications in the phase space structure accompanying the departure from integrability.
The investigation of these issues and of their consequences is however beyond the scope 
of these pages, and will be discussed elsewhere.

To proceed towards our goal, {\it i.e.}, a Thermodynamics for SGS, in the figures 
\ref{fig:FPULyap} and \ref{fig:NBLyap},
we compare the energy dependence\footnote{To be precise,
for the {\sf FPU} model, the perturbation parameter is the product $\beta\varepsilon$, 
where $\beta$ is the anharmonicity constant.} of the maximal Lyapunov exponent, computed
either using the standard algorithm \cite{BGS} or the analytic geometrodynamic
formula, equation (\ref{eq:LCN}). 

Figure \ref{fig:FPULyap} displays the transition in the scaling behaviour of $\gamma_1(\beta\varepsilon)$
outlined above.

In figure \ref{fig:NBLyap} are reported the values
of the maximal Lyapunov exponent as a function of the specific energy of the system, for two series
of simulations with $N=400$ and $N=4000$ bodies. It is evident the power-law dependence
of $\lambda_1(\varepsilon)\propto |\varepsilon |^{3/2}$. This dependence however
disappears if we measure the instability exponent in the unit of the {\sl dynamical time} $t_D$, which
is customarily defined for SGS as proportional to the inverse square root of the density
$t_D\propto (G\rho)^{-1/2}$. So, for SGS, we find that the adimensional chaoticity parameter
$\gamma_1$ is constant and does not depends on the amount of binding energy in the system.

\begin{figure}
%\seteps{0cm}{12cm}{10cm}{FPULyap.eps}
\caption[]{Scaling of the dimensionless chaoticity indicator $\gamma_1=\lambda_1\, t_D$ with
the perturbation parameter $\beta\varepsilon$ for a FPU chain, as obtained
according to the analytic formula, eq.(\ref{eq:LCN}). Open circles refer to the values
computed  from  mean values and fluctuations obtained from phase space averages \cite{MarcoLapoLivi},
whereas, starlike symbols with error bars show the values obtained from time 
averages computed along numerically integrated trajectories \cite{CPTD}.
}
\label{fig:FPULyap}
\end{figure}

From the comparison of the two figures the peculiarity of SGS emerges clearly: 
{\sl they behave, at any binding (specific) energy, as standard 
many body systems above the {\it strong stochasticity threshold}}.

\begin{figure} 
%\seteps{0cm}{12cm}{10cm}{NBLyap.eps}
\caption[]{Scaling with the binding energy per particle, $\varepsilon$, 
of the standard maximal Lyapunov exponent $\lambda_1$ and
of the dimensionless chaoticity indicator $\gamma_1=\lambda_1\, t_D$  for two sets of gravitational
N-body systems, with $N=400$ (circles) and $N=4000$ (stars) bodies, respectively.
}
\label{fig:NBLyap}
\end{figure}

This not only confirms that bound collisionless SGS are characterized by a dynamical
instability proceeding at a very fast rate\footnote{This is indeed a very well known fact:
it was discovered already by Miller in the '60s, when he performed the first pioneering numerical
simulations of N-body gravitational systems and has been confirmed many times also in the
last years.}, but also suggests that they possess the 
strongest statistical properties, in analogy with those of {\sl standard} dynamical systems
in the regime of fully developed stochasticity.

Once realized this, we are allowed to invoke the {\it Chaotic hypotesis} outlined in
the Introduction, and to build a Statistical Mechanics for SGS.

\section{Chaos and Statistical properties of Dynamics. \label{sec:FromDynTOSM}}
If the dynamics of a MDoF system possess strong enough stochastic properties, the Chaotic
hypotesis \cite{GC95} asserts that the averages of phase space functions can be computed
as if the system were a transitive Anosov flow. From a physical viewpoint, this amounts
to say that the system attains the most probable macroscopic equilibrium state compatible
with the global constraints imposed at the initial time\footnote{These constraints are
essentially the values of the possible conserved quantities, like energy, linear and angular
momentum and so on.}.

So, if we are able to find this equilibrium state, we can compute all the averages using
the invariant measure over the stationary distribution. The analogy with Anosov systems
says even more, it gives also an indication about the {\sl rates} governing the approach
to the equilibrium state, which should be exponential if the manifold is {\it compact}.

For most MDoF systems these equilibrium (or stationary non equilibrium) states coincide either with
the canonical or microcanonical ensemble, according to the system being in contact with a thermal
bath or isolated, respectively.
If the interaction potential governing the internal dynamics satisfies the 
conditions outlined in the Introduction, then the theorems of Statistical Mechanics entail that  
the descriptions based on the two ensembles agree, apart from terms which vanish as $1/N$ in 
the {\sl thermodynamic limit}, {\it i.e.}, when the number of particles, the spatial volume
and the energy of the system grow indefinitely, keeping however constant their ratios:
\be
N\rightarrow\infty\ ;\ \ \ V\rightarrow\infty\ ;\ \ {\cal E}\rightarrow\infty
\ee
with 
\be
n\doteq \frac{N}{V} =constant\ ;\ \ \ \varepsilon\doteq\frac{\cal E}{N}=constant\ .
\ee

\subsection{{\it Obstacles} to a Statistical Mechanics for gravitating systems.}
From the above discussion, one amongst the many difficulties for a Statistical Mechanics of 
{\sl gravitationally bound} SGS emerges immediately: it is not possible to perform the 
thermodynamic limit complying with the above requirements. 

From the expression of the gravitational N-body Hamiltonian,
equation (\ref{eq:NBHam}), it is immediate to realize that keeping constant the density
$n$ implies that the specific energy grows without limit, $\varepsilon\propto N^{2/3}$.
This problem derives obviously from the fact that, for {\sl bound} SGS, the number of
bodies (or better, the total mass), the energy, and the {\sl spatial volume} cannot
be fixed independently from each other, as it is instead possible for {\sl standard}
laboratory systems. 

It is usually stated that that the one just mentioned is only the most evident difficulty and that
another one, in principle not less relevant,
is related to the fact that {\sl the most probable state} for the gravitational
N-body system cannot be determined, because {\it formally}, the measure of the phase 
space volume accessible for a given total energy is virtually infinite.

However, if we are not interested in the properties of the abstract, eternal, {\it final}
equilibrium state, but rather to determine the quasi stationary states whose secular
evolution proceeds on time scales much longer than any dynamical process\footnote{And, from
the astrophysical viewpoint, also much longer than the lifetimes of stellar systems to which
we plan to apply at the end all these theoretical considerations.},
the second of the above mentioned difficulties can be surmounted, as we will show, in a
(hopefully) self consistent way. 

The remarkable fact is that, to overcome this obstacle,
we do not need to avoid or mask the peculiarities originating it, but rather we succed
to incorporate their implications, to outline a hierarchical evolutive path, able to describe 
the succession of quasi-equilibria of stellar systems, covering a range of timescales 
spanning from those of the order of the virialization (or Violent Relaxation) up to 
those related to the formation of a collapsed-core--expanding-halo structure. 
As long as this hierarchy of timescales is covered, dynamical processes involving
finer and finer spatial scales take place and this evolution is actually endless, as the
{\sl final} equilibrium state can be attained only on infinite times, and just for this,
does not bear any direct relevance to {\sl real} astronomical objects, once verified
that the secular evolution of the metastable states proceeds on timescales longer than
the stellar systems lifetime.

From the above discussion, it follows that the sole serious difficulty for a {\sl standard}
statistical description of N-body SGS is related to the non existence of the {\sl standard}
Thermodynamic limit. As we will discuss below, this is not a real hindrance for a
self consistent Thermodynamics of SGS, as it is well known that many other {\sl laboratory}
systems\footnote{The so called {\it "small systems"}, see below.},
characterized by non-extensive interactions, do not possess a well defined Thermodynamic
limit\footnote{
Rather, the point is that, loosely speaking, the behaviour predicted in the T.L. does not
bear any resemblance with the phenomenology observed in real, finite size, systems. And this should
not to be considered surprising, because, for those systems, some {\sl peculiarities} come just from 
the finiteness and disappear when the size exceeds the range of the interaction.} 
and, nevertheless their macroscopic or collective behaviour can be described coherently
using a suitable thermodynamic formalism which take into account their peculiarities.

The analysis below shows that a conceptually coherent framework can be settled to describe
also SGS, certainly intrinsically even more peculiar than  the {\sl small systems}
outlined above, but nevertheless suitable for a self consistent Thermodynamics.  

\subsection{Dynamical singularities and Statistical Mechanics.}
The above scenario owns also a dynamical counterpart which can be easily understood within the
Geometrodynamical framework. Without going too much into details, we remark that the convergence
of time averages to phase space values of geometric quantities ({\tt e.g.}, the curvature)
proceeds in the case of SGS on timescales which strongly depend on the spatial resolution
over which we compare the averages themselves. That is, the time required for the time averages
(and fluctuations around them) to converge to the {\sl static} averages is very sensitive
to the spatial resolution used to compute the latter. Incidentally, we remark that the 
ergodicity time for the {\sl Ricci curvature} is infinite if the mathematical
definition of the Newtonian interaction is assumed. It becomes viceversa a well defined,
finite, quantity, if the physical form of the interaction is considered. It must be stressed
that even in this respect, as the spatial resolution increases, the ergodicity timescales 
of geometrodynamic quantities grow up, exactly at the same rate of the times associated 
to the relaxation of statistical quantities.

\subsubsection{Mathematical vs. Physical peculiarities of gravitational interaction.}
We conclude this section listing  the more relevant peculiarites of the Newtonian
interaction outlining also which, if any, amongst them could constitute a real obstacle to a 
functional statistical mechanical foundation of a Thermodynamics for SGS.
\begin{description}
\item{\it Singular interaction at short distances:}
The most evident singularity of the gravitational interaction is obviously related to the
divergency of the potential when two particle get closer and closer. Although clearly
of outmost importance for the single particle dynamics  and for the evolution of {\it few 
body} systems, the global and statistical relevance of close encounters is practically negligible
over any finite timescale, as the frequency of high energetic close encounters is vanishingly small and
the effect on the global quantities irrelevant, compared with that of the overall dynamics. 
Then, even if it is physically plausible\footnote{And numerically very useful, especially if low 
order efficient symplectic integrators are used, in order to guarantee the faithful preservation 
of the Hamiltonian properties of the flow.}, from a statistical viewpoint there is no need of any 
{\it softening} of the potential.
\item{\it Non confining nature of the potential and divergence of the phase-space volume:}
The configuration space of a N-body bound SGS  obviously diverges, and in principle very badly, indeed
as 
\[ \Gamma_{\bf q} \sim \lim_{V\rightarrow\infty } V^{N-2}\ .\]
 This is is due to the fact
that Newtonian interaction is unable to confine, asymptotically,
inside a finite volume more than two bodies.
If the mathematical N-body system is considered, an analogous divergence comes out also from
the volume in the momentum space. However, as long as $\log{N}\gg 1$, the characteristic time
for the evaporation is considerably longer than any dynamical timescale, and the evolution is only 
secularly influenced by the slow evaporation of stars, which, incidentally,
 causes the remaining system being more and more bound, so inhibiting further escapes.
\item{\it Unscreened, long-range nature of the interaction:}
It is the singularity which most affects the statistical and thermodynamics properties of
SGS. This is in part obvious as it concerns the dynamics over all the spatial scales, and consequently,
according to the general discussion above, every timescale. The long range nature of the interaction
reflects on the efficiency of collective phenomena, among which the Violent Relaxation is surely the most
popular but it is not the only one. The absence of any effective screening mechanism entails that
gravitationally bound systems do not own any characteristic lenght, and reflects on the scale invariance
of most features. From  statistical and thermodynamical viewpoints the most far reaching consequences
reflect in the following points:
\begin{itemize}
\item N-body systems are not {\it stable} in the sense of {\sl Rigorous Statistical 
Mechanics} \cite{Ruelle}: given any number of particles $N$, the binding energy per particle can increase 
without any upper bound ({\it i.e.} the energy does not have a {\it lower} bound).
In particular, the binding energy of the {\it fundamental state} of a system scales at least as $N^2$.
That the lack of stability property is related to the long range nature of the potential
and it is unaffected from the singularity at short distances emerges clearly from the discussion above 
and from many numerical simulations performed recently \cite{MCSCPMP00}.
 Indeed, the introduction of a softening of the potential is able
to introduce an upper limit for the binding energy of the system, but is unable to {\sl cure}
the lack of the {\it stability} property\footnote{Even the most {\sl radical} modification of the short range
interaction, {\it i.e.}, the introduction of an hard-core repulsive potential at short distances,
does not prevent the divergence of the binding energy per particle in the thermodynamic limit, though being
able to make it finite at any finite $N$.}.
\item From the previous point, it follows that the internal energy of a system scales as $N^2$, and this
remains true for most thermodynamic potentials. As the extensivity of the latter is a necessary
condition for the proof of the ensemble equivalence, the long range nature of the Newtonian interaction
is the main cause of the peculiarity of Gravitational Thermodynamics.
\end{itemize}
\end{description}
From the above discussion it emerges clearly that is the non-extensivity of potential energy
(and, by the virial theorem, also of the kinetic and total energies) to imply the non equivalence
of statistical ensembles, and this requires, in turn, a deeper analysis in order to understand which
ensemble, if any, is best suited to describe SGS.

\section{From Statistical Mechanics to Microcanonical Thermodynamics 
of self gravitating systems. \label{sec:TD}}
Up to now we worried about the {\sl rigorous theorems} concerning the equivalence of ensembles,
the scaling of macroscopic quantities in the thermodynamic limit and so on. However, as emphasized
in previous sections, in recent years it has realized that the issue of the inequivalence
of statistical ensembles emerges whenever the scale of the dominating interaction is comparable with
the spatial extension of the system under study. 
It is easy to understand that in such circumstances the interaction energy ceases to be an extensive
quantity and, if some {\sl equipartion property} holds, the same is true also for  kinetic and
total energies.
However, many of these so-called {\it small systems} \cite{smallsystems} possess a well defined, 
though somewhat peculiar, thermodynamic behaviour. 
In these systems it is not difficult to encounter regions in the
space of thermodynamic parameters where the specific heats become negative. When described within the
microcanonical ensemble, all these peculiar phenomena do not originate any {\sl paradox}, which appears
instead as soon as a {\sl canonical ensemble} approach is attempted. The best that can be done,
sometimes, within this framework, is to describe the phenomenology inside these regions as a 
phase transition, though {\it true} phase transitions themselves are {\sl rigorously} defined 
only in the Thermodynamic limit, which we know to be a rather meaningless limit just when the system
properties are not extensive.

\subsection{Ensemble inequivalence.}
As discussed in the previous sections, the non extensive nature of the energy entails, among others,
the inequivalence of statistical ensembles. In order to find out what among the classical ensembles is 
best suited to describe SGS we first observe that, if we consider {\sl isolated} N-body systems, nothing
suggests to suppose them in contact with any heat bath. Going a little more below the surface, we
observe more concretely that if we require, according to Boltzmann, that the ensemble fulfils the
{\it orthodic} property (see \cite{GallaMS}, \S 5), then neither the canonical nor the grand-canonical
ensembles can be used to describe gravitationally bound N-body systems. Heuristically,
this simply because both
these ensembles fix the {\sl Temperature} as a constant parameter of the distribution, assuming
implicitly that this quantity is constant throughout every system of the ensemble. 

To this somewhat formal point, we have to add the already quoted, and physically most relevant, 
incongruence met by the canonical (and also grand-canonical) approach: 
SGS are characterized by negative heat capacities, whereas
equation (\ref{eq:CV}) shows that within the canonical ensemble the heat capacity is positive
by definition.
Moreover, the non extensivity of potential, kinetic and total energies entails the analogous
non extensivity of canonical and grand-canonical thermodynamics potentials, as the Helmholtz
free energy $F$ or the Gibbs potential $G$.  These features not only imply the non equivalence
of ensembles but also indicate that even the definition of averages (and even more of fluctuations)
of thermodynamic variables becomes problematic.\\

Up to this point we analyzed the problems affecting the canonical and grandcanonical descriptions.
Now we turn to ask how the peculiarities of gravitational interaction reflect also
on the microcanonical description.

Before to proceed, we think that this seems to be the appropriate point to recall a general
statement: the {\sl strong} chaotic properties of dynamics, the Poincar\'e-Fermi theorem,
the general theorems of Ergodic Theory, all focus on the relationships between the stochastic properties
of motions and the occurrence of a uniform diffusion of initially concentrated phase space
volumes over the whole region accessible to motion, compatibly with the integrals of motion
allowed by the system dynamics. If such a spreading process in phase space is efficient enough,
then, those general theorems, also according to the {\it Chaotic Hypotesis}, assert that the
the time averages of dynamical variables coincide with the phase averages computed over the
region of phase space where motions take place transitively. If we are dealing with  Hamiltonian
systems, this region is the constant energy surface\footnote{Possibly taking also into account
the conservation of other global integrals, like the position and linear momentum of the center
of mass of the system and the total angular momentum if only central forces are present. This
reduces the dimensionality of $\Sigma_{\cal E}$ from $6N-1$ to $6N-10$, which, for $\log{N}\gg 1$,
does not modify obviously the macroscopic behaviour.} $\Sigma_{\cal E}$. This amounts to say that
the {\sl strong enough} stochastic properties of Dynamics of MDoF Hamiltonians, naturally lead
to a {\sl microcanonical} statistical description. 

The task of the powerful formal apparatus
of {\sl rigorous} Statistical Mechanics is then {\sl to prove} the equivalence of ensembles. In case
of success, then we can use the more {\sl tractable} expression of canonical or grand-canonical
ensembles to describe the macroscopic properties and the Thermodynamics of our system. If the 
equivalence does not hold, then the sole Statistical description allowed is the one compatible
with the global constraints on our system: if it is isolated, then only a microcanonical
description is {\sl legitimate}.\\

From the above discussion, it is clear that the use of {\sl microcanonical description} is 
justified, provided that no internal inconsistencies occur; and that the inequivalence of
ensembles, if ever, should suggest a critical reconsideration of the results obtained
using a canonical or grand-canonical description.

It is easy to see that the only trouble comes from the formal unboundness of the phase space
volume accessible at a given constant energy. Nevertheless, if we are interested in the
finite time quasi equilibrium states, we can take the argumentations discussed in the previous 
sections to emphasize that, at any finite time the {\it effective} phase space volume accessible
is well defined in terms of the global quantities describing the state of the system (energy per
particle, specific angular momentum, etc.) and it is finite\footnote{It is easy to verify that even
in the presence of escaping particles or taking into account the occurrence of close encounters, the
contribution of these events to the actual phase-space volume is negligible as long as $\log{N}\gg1$
for times of  $t\ge {\cal O}(Nt_D)$.}.

So, for any finite scale of time, we are able to derive an {\it effective} expression for the 
microcanonical entropy, secularly evolving, and able to take into account the increasing effects 
of the phenomena occurring on smaller and smaller spatial scales, and, correspondingly,
on longer timescales.\\

Starting from the Hamiltonian of the system (\ref{eq:NBHam}), given a value for the total energy
of a {\it bound} SGS\footnote{
It is possible to perform a statistical mechanical treatment even for loosely bound systems or 
systems in interaction with some background gravitational field. This leads to a very interesting
phenomenology, indicating the occurrence of critical phenomena in 3D-gravitational systems. The details
of this work are described in \cite{MCSCPMP00}.
}, 
${\cal E}<0$, we can define a set of {\sl macroscopically averaged} quantities:
\begin{description}
\item{\it the energy per particle:}
\be
\varepsilon\doteq \frac{\cal E}{N} \le 0\ ,
\ee
\item{\it the virial ratio:}
\be
{\cal Q}(t) \doteq \frac{-2 K(t)}{U(t)}\equiv 
\frac{\sum\limits_{i=1}^N {\frac{p_i^2}{m_i}}} {\frac{1}{2}\sum\limits_{i=1}^N\sum\limits_{j\ne i} 
{\frac{m_i m_j}{r_{ij}}}} \ ,
\ee
where the indication of the time dependence of the kinetic and potential energies, $K(t)$ and $U(t)$, 
respectively, is a shortcut to recall that their values depend on the state of the system at time $t$;
\item{\it the inertial radius:}
\be
R_i(t)\doteq \left[ \frac{\sum\limits_{i=1}^N m_i r_i^2}{\sum\limits_{i=1}^N m_i}\right]^{1/2}\ ,
\ee
\item{\it the harmonic (or {\it virial}) radius:}
\be
R_g(t)\doteq \frac{\left( \sum\limits_{i=1}^N m_i\right)^2}
{\sum\limits_{i=1}^N \sum\limits_{j\neq i} \frac{m_i m_j}{r_{ij}}}\ ,
\ee
\item{and the ratio among them:}
\be
\alpha(t)\doteq \frac{R_i(t)}{R_g(t)}
\ee
\end{description}

In the simplifying hypotesis that the mass spectrum is nearly flat, so that the average
mass $\bar{m}\doteq\langle m_i\rangle_i$ represents the typical mass of the system\footnote{When 
a very steep spectrum of masses is allowed, the computations are only a little more involved, but the results
presented below are unchanged. In addition it is possible to obtain also the dependence of the entropy
on the mass sectrum.}, it is a simple computation to obtain an expression for the microcanonical
entropy of the system, which reads
\be
S\equiv k_B \,\sigma(N,\varepsilon,{\cal Q},\alpha,\bar{m})\ ,
\ee
where 
\be
\sigma = \frac {3N}{2} \ln\left[ \left( {\frac{G^2N^2\bar{m}^5}{-2\varepsilon}}\right) \, 
\alpha^2(t)\, {\cal Q}(t)\, [2-{\cal Q}(t)]\right]
\label{eq:CPentropy}
\ee
and it is obviously understood that the also the varying quantities $\alpha(t)$ and 
${\cal Q}(t)$ depends on time only through the actual state of the system.\\

The above argument can be extended to take into account other details about the state of the system,
but the simplest derivation sketched here is sufficient to emphasize a number of relevant
points.

It is evident that the microcanonical entropy (\ref{eq:CPentropy}) gives correctly  the equilibrium
conditions related to variations of the parameters on which it depends.

However, the most striking result is that it turns out to be {\it an extensive quantity}, being proportional
to the number of particles\footnote{
A very general approach to the definition of suitable microcanonical entropies for 
generic MDoF systems is discussed in \cite{CPentropy}.
}.

Another very important feature of this entropy is that it is possible to derive from
it easily all the relevant thermodynamic quantities, which agree with the phenomenological
expressions obtained at a rather heuristic level.

The most important is the {\sl microcanonical temperature}
which, coherently with the statistical description,
 is clearly only an average quantity for the systems of the microcanonical ensemble\footnote{
According to what stated above, in the derivation of the microcanonical thermodynamic
relations, we take into account that, for bound SGS, the energy, the number of bodies and
the volume are not independent quantities.}
\be
\frac{1}{T}\doteq \left(\frac{\partial S}{\partial {\cal E}}\right)_{N} \equiv \frac{1}{N} 
\left(\frac{\partial S}{\partial\varepsilon}\right)_{N}
\ee
which, using (\ref{eq:CPentropy}), gives, coherently with the fact that bound systems have 
negative energy and positive temperature,
\be
\varepsilon = -\frac{3}{2} k_B T\ .
\label{eq:Tdieps}
\ee
Consequently, a negative heat capacity follows:
\be
C_V \doteq \left(\frac{\partial {\cal E}}{\partial T}\right)_{N} = - \frac{3}{2} N k_B \ ;
\ee
that is, a {\sl negative specific heat}, $c_V = -3/2 k_B$.\\

The surprising features of the entropy (\ref{eq:CPentropy}), defined in such a straightforward way,  do not limit
to single out the thermodynamic properties of the system, but allows also to determine the
dynamical evolution of the global quantities.
It is immediate to see, for example, that, as a function of the virial ratio,  
the entropy (\ref{eq:CPentropy}) {\it is maximized when} ${\cal Q}=1$:
\be
\frac{\partial\sigma}{\partial {\cal Q}} = \frac{2(1-{\cal Q})}{{\cal Q}(2-{\cal Q})}
\ee
That is, during the approach towards the virial equilibrium
the average entropy of the system increases to the maximum value allowed by the values of
the other parameters. Equivalently, it can be said that, starting from an initial state out of
the virial equilibrium, entails a rapid relaxation to the maximum entropy state compatible with
the given energy.
That most of the volume of phase space of N-body systems is occupied by configurations in virial equilibrium
is evident from figure \ref{fig:SdiQ}, where the probability distribution of the ${\cal Q}$ values
is shown to be strongly peaked around the equilibrium value already for moderately large N-values.

 \begin{figure} 
%\seteps{1cm}{10cm}{8cm}{SdiQ.eps}
\caption[]{Probability distribution for the values of the virial ratio, determined according to the 
definition of the microcanonical entropy, equation (\ref{eq:CPentropy}), 
for bound self gravitating systems with moderately large number of bodies: from the broader to the
narrower distributions it is $N=10,\, 50,\, 200,\, 1000$. It is well evident that, as the number of bodies
increases, the overwhelming fraction of $\Sigma_{\cal E}$ is occupied by systems in virial
equilibrium.}
\label{fig:SdiQ}
\end{figure}
We think that it is very impressive to see how such a definition of {\sl finite time entropy} is able to
predict even the dynamical states maximizing locally (in time) the entropy.

As a further evidence of this property, we observe that it is also possible to predict 
(and also to numerically verify), that the secular dynamical evolution
of a gravitationally bound system leads the ratio $\alpha(t)$ to an average steady growth, over which
small quasi periodic modulations are imposed. The expression derived above for the microcanonical entropy
correctly describes also this secular dynamical evolution through locally stationary states
of slowly increasing entropy! 
\be
\frac{\partial\sigma}{\partial\alpha} \propto  \frac{N}{\alpha} > 0\ .
\ee
This means that the entropy will increase along with the growth of the parameter $\alpha(t)$
(because, supposing fixed all the other parameters, we have $\dot{\sigma}\propto N\dot{\alpha}/\alpha$).
Well, it is easy to see, even analytically, that the development of a core-halo structure
implies an increase of $\alpha(t)$. The relationship between $\sigma$ and $\alpha$ however indicates
that the entropy growth rate, though remaining positive,  slows down as $\alpha$ steadily
increases.

Adding a further comment to the results above, we remark that, while the apparent singularity 
of the entropy when  $\varepsilon\rightarrow 0^{-}$ is fictitious, as, in this limit 
it is also ${\cal Q}\rightarrow 2^{-}$, the singularity for ${\cal Q}\rightarrow 0^{+}$ is instead
real, and again very enlightening on the sensible meaning of the microcanonical entropy here
presented. Indeed the above limit describes a {\sl completely frozen} state of the system, 
that is, a given configuration with all the bodies having zero velocity. The probability
of such a state is zero for any $\varepsilon < 0$, and correctly its entropy diverges.
Incidentally we notice that, if $\varepsilon\rightarrow 0^{-}$, then a {\sl frozen} state
become possible, and accordingly $\sigma$ can be made finite.\\

We emphasize again how the present definition of microcanonical entropy is  effective in taking into
account the secular evolution of the system, through the full hierarchy of timescales associated
to the singular dynamical processes occurring in the asymptotic evolution of SGS. It is also remarkable
that, if we consider a somewhat {\it regularized} gravitational N-body system, where the divergence 
at short distance is removed ({\tt e.g.}, through a usual softening procedure, for example) and
the unboundness of configuration space accessible is avoided, introducing a kind of {\sl cosmic vessel},
then the microcanonical entropy will still entail the secular evolution, but will be in this
case able to single out an {\it end state} that is now well defined and attainable in a finite time! 
We stress that this is not at all a trivial result, because
the {\sl regularization} procedure described above does not modify the statitistical and thermodynamic
peculiarities of the system, which remains nonextensive and for which the inequivalence of
ensembles remains true, as long as the system is gravitationally bound. In particular, such a
regularized system will keep a negative heat capacity in a wide range of specific energies and
the canonical ensemble description permains unable to grasp even the qualitative features
of its Thermodynamics \cite{MCSCPMP00}!

\section{Comments and Conclusions. \label{sec:Concl}}
We presented an {\sl effective} {\it microcanonical entropy} able  to describe correctly
not only the thermodynamic properties os SGS, but also to outline the sequence
of secular evolutionary processes leading the systems towards states of higher and higher
entropy.

The dynamical and statistical justifications of a microcanonical description of gravitationally
bound N-body systems have been derived and discussed on the grounds of analytical and numerical
evidences of their {\sl strongly chaotic properties}, which allow the assumption of 
validity of the {\it Chaotic Hypotesis} of Gallavotti and Cohen. The results obtained
within the {\sl Geometrodynamical Approach to Chaos} gave further support to the 
reliability of this assumption.

The peculiar dynamical properties of gravitational interaction, 
the breakdown of the extensivity and the consequent inequivalence of classical
statistical ensembles, moreover, rule out any reliable canonical or grand-canonical description 
of the Statistical Mechanics and Thermodynamics of {\it gravitationally bound} N-body systems. 

It can been shown that the microcanonical entropy is the sole thermodynamic potential which
keeps the extensivity at any finite time even for the non stable Newtonian interaction potential,
while all thermodynamic potentials defined within the canonical or grand-canonical ensembles
lose the extensivity property and consequently lead to unclear, unuseful or wrong
thermodynamic relations.
A simple sketch of the above assertions follows directly from the definitions of the canonical
and grand-canonical potentials: 
\begin{description}
\item{\it Helmholtz free energy.}\\
 In the canonical ensemble, given the internal energy 
\be
U\equiv\langle H\rangle_\beta \cong {\cal E}\ ,
\ee
where average is performed on the canonical ensemble and ${\cal E}$ is instead the fixed energy
of the {\sl corresponding microcanonical ensemble}\footnote{We are using the relations which are
valid when there is the ensemble equivalence. Here we use them to further demonstrate the
difficulties of canonical description!}; the Helmholtz free energy is related
to it by the relation
\be
F = U - T S \ .
\ee
From the results obtained above, we see clearly that, for a non extensive system
(be it a SGS or a {\it small system})  it is $ F\propto N^2$; that is, $F$ is not
more an extensive quantity, although $S$ it is\footnote{At this point we notice that the above
derived expression for the temperature, equation (\ref{eq:Tdieps}), gives, 
$T\propto\varepsilon$. For non extensive systems this entails that also the product 
$T\,S$ is proportional to $N^2$, coherently with all the above discussions.}.
\item{\it Gibbs potential.}\\
 In the grand-canonical ensemble,  the appropriate potential is given
by the Gibbs function, whose thermodynamic relationship with $F$ is given by
\[
G = F + PV
\] 
and, as it is easy to verify, both the terms of the sum are, for a {\sl bound} SGS, proprotional
to $N^2$. 
\end{description}

If a {\sl physically reasonable regularization} of gravitational interaction, which
does not modify its characteristic peculiarities, like the non extensivity and the
{\it non stability}, is introduced, the proposed microcanonical entropy keeps completely
its validity and moreover becomes able to single out not only the hierarchy of secularly
evolving meta-equilibrium states, in analogy with the non regolarized case, but also to
determine a non singular {\it final}  equilibrium state, which was only asymptotically
reachable prior the regolarization.

In addition, the above regularization allows for a detailed study of the critical
phenomena occurring in SGS \cite{MCSCPMP00}, able to show the occurrence of phase transitions
in gravitational N-body systems and also to confirm the doubtful reliability
of a canonical description of the Statistical Mechanics and Thermodynamics of
Gravity .\\

The above discussion points out that the peculiarity of the collective properties
of gravitational systems depend largely on the long range attractive nature of the interaction
and it is almost completely independent on the formal singular behaviour at short distances.
Taking into account that the latter is only a formal singularity, we can safely ignore it
also on the grounds of the results showing its complete irrelevance for the statistical
properties of SGS. The details on the way of coping with the dynamical consequences of the presence of the
singularity affect the evolution of macroscopic quantities only on timescale which are
longer than the binary relaxation time, in some cases even exponentially larger! 

\subsection{Further comments and future developments.}
A further point deserves to be remarked: it is undeniable that the {\it limits} on which most 
of the {\sl standard} Statistical Mechanical and Thermodynamic frameworks rely are of
outmost relevance for the comprehension of the behaviour of most macroscopic systems. For the 
overwhelming largest class of MDoF systems, the thermodynamic limit allows to understand the basic
and usual properties of macroscopic matter and even the rich and complex phenomenology of critical
phenomena. At the same time, theorems assuring the existence of an invariant distribution
are exceptionally important to single out the properties of the final equilibrium states of most
macroscopic systems. 

Nevertheless, it is also evident from the considerations here presented, that the operation of taking
these limits ({\it i.e.}, the thermodynamic limit and the infinite time limit),
assigning to them the role of selection criteria in order to establish whether {\sl a Thermodynamics}
for a given class of systems can be constructed or not, can hinder in some cases
the possibility of obtaining alternative statistical and thermodynamic description of MDoF systems 
which are certainly {\sl singular} in some respect, but possess notwithstanding well defined, though peculiar,
thermodynamic properties, in general for many of them, or at least over any physically reasonable
finite interval of time, as it happens for SGS.\\

The outcomes presented above are enlightening also for the critical analysis of the results 
based in some way on the assumption of a canonical equilibrium of gravitational N-body systems. 
In particular, we claim that many results based on the maximization of entropy like quantities constructed
using the single particle distribution function have to be reconsidered.\\

In conclusion, despite the {\sl rigorous theorems}, we have shown that a thermodynamic
description of gravitational N-body systems is not only feasible, but also suggests a 
new approach to {\sl peculiar systems} for which the standard paradigms of {\sl rigorous}
Statistical Mechanics do not apply. 

We plan to further develop the line of research
here proposed in order to make it possible to obtain indications on the detailed
evolution of SGS on the basis of thermodynamic arguments alone. 

In order to attain
this goal it is necessary to have more informations on the detailed energy distribution
of N-body systems and for this the use of extensive numerical simulations is required.

\acknowledgements
The work of P.C. is partially supported by {\sf C.S.S.} under the initiative {n.2000A:MPDS}.

\end{article}
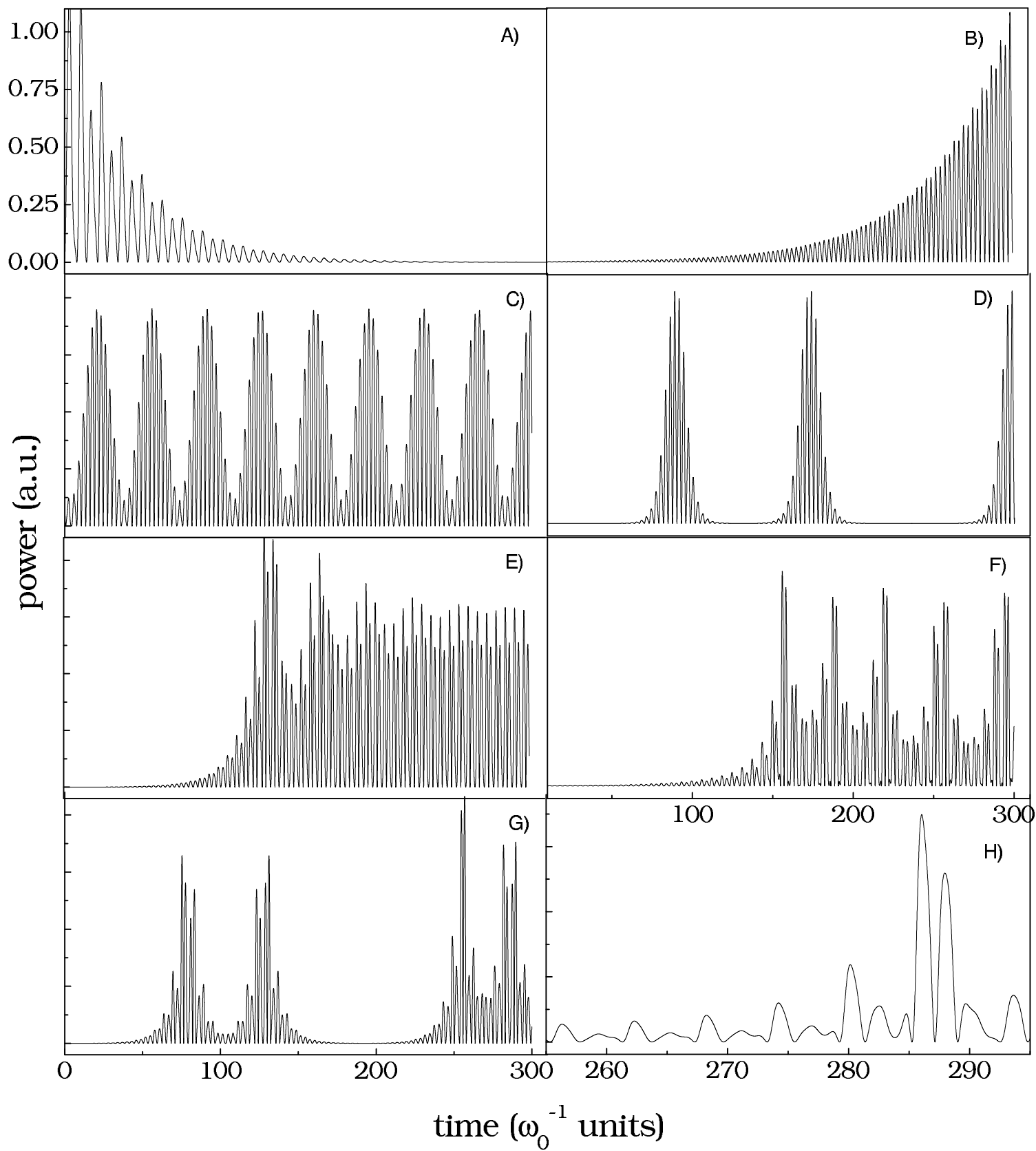
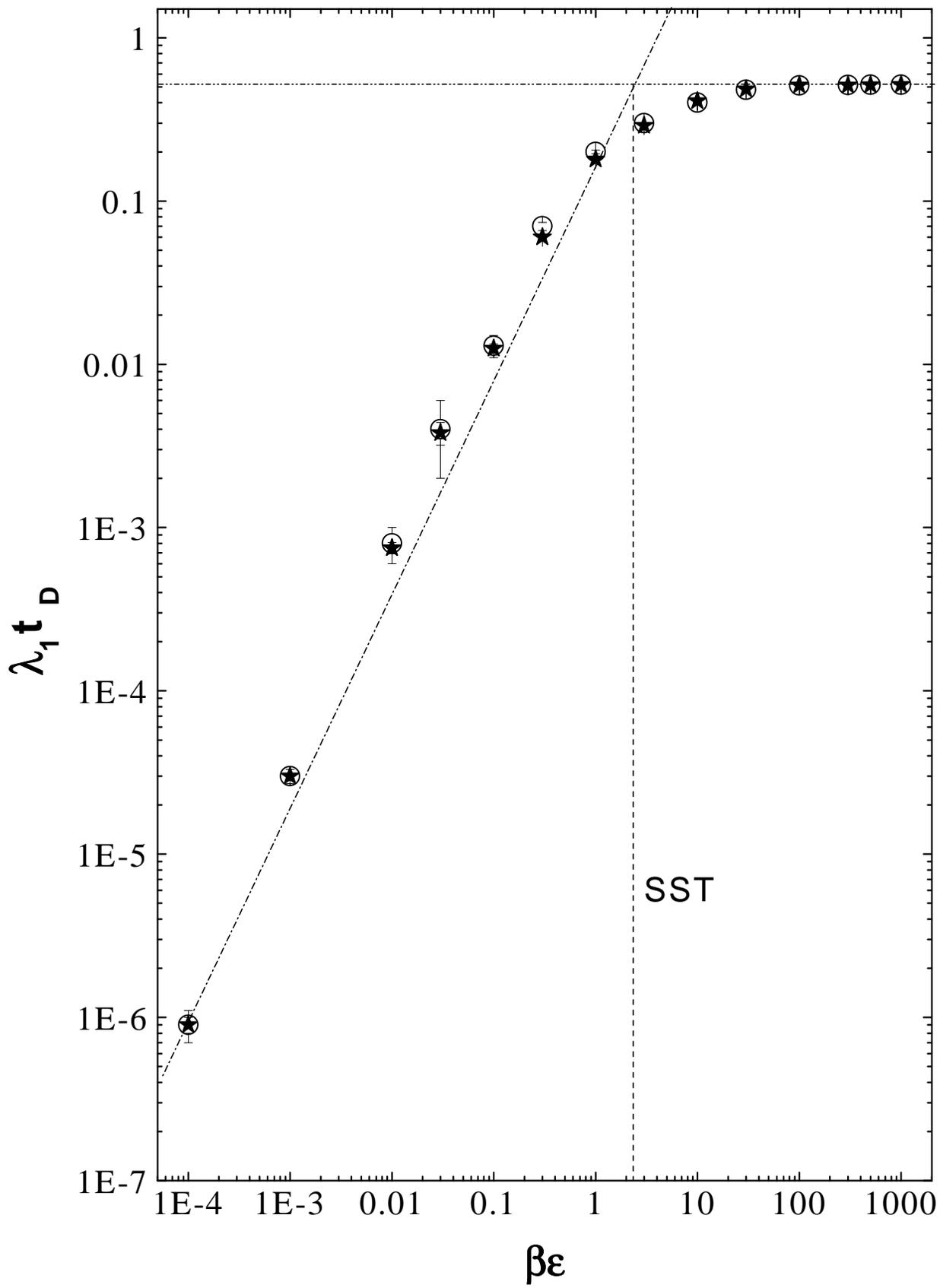
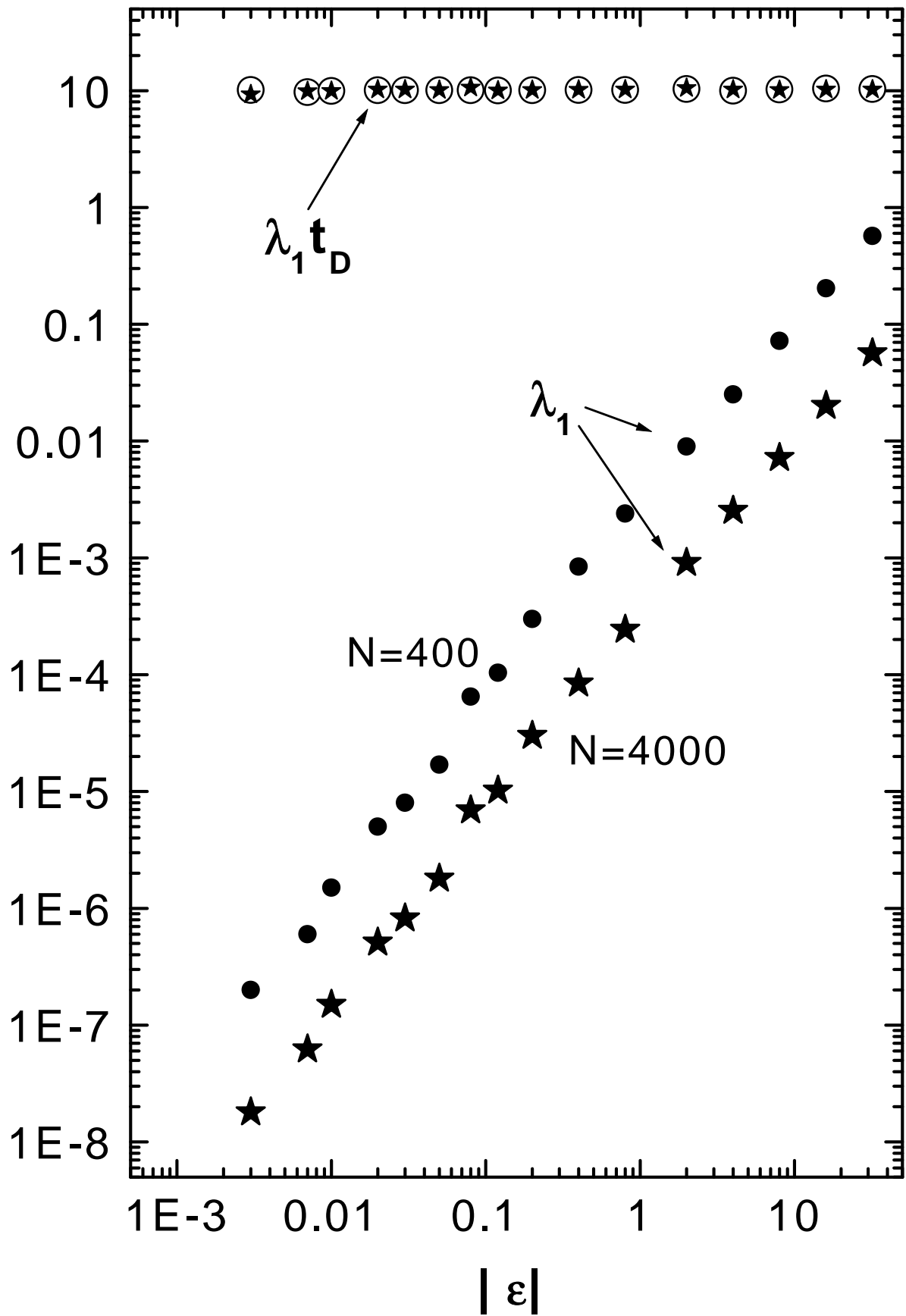
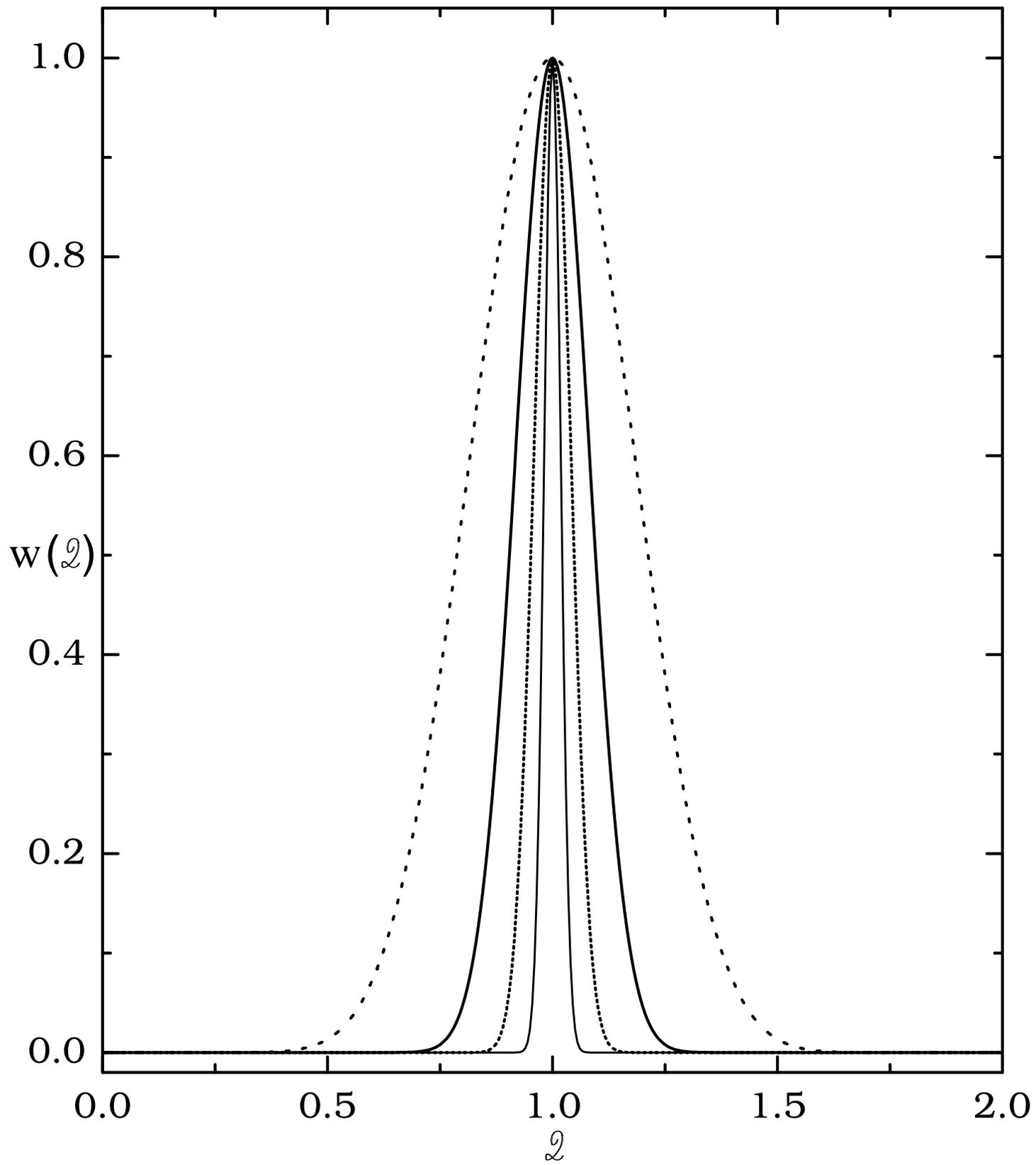
\end{document}